\documentclass[usenatbib]{mn2e}

\usepackage{epsfig,amsmath,graphics}

\newcommand{\aap}{A\&A}

\newcommand{\aj}{AJ}
\newcommand{\apj}{ApJ}

\newcommand{\mnras}{MNRAS}
\newcommand{\pasp}{PASP}

\title[Effect of metallicity on Cepheid magnitudes in M33 ]{The effect of metallicity on Cepheid magnitudes and
  the distance to M33\thanks{Based on observations obtained with the
    WIYN 3.5m telescope, Kitt Peak National Observatory, National
    Optical Astronomy Observatories, which is operated by the
    Association of Universities for Research in Astronomy, Inc. (AURA)
    under cooperative agreement with the National Science
    Foundation. The WIYN Observatory is a joint facility of the
    University of Wisconsin-Madison, Indiana University, Yale
    University, and the National Optical Astronomy Observatories.}}

\author[V. Scowcroft et al.]{V.~Scowcroft,$^1$\thanks{Email:
    vs@astro.livjm.ac.uk} D.~Bersier,$^1$ J.R.~Mould,$^2$ and P.R.~Wood$^3$ \\
$^{1}$Astrophysics Research Institute, Liverpool John Moores
University, Twelve Quays House, Egerton Wharf, Birkenhead, CH41 1LD,
UK\\
$^{2}$School of Physics, University of Melbourne, Parkville, VIC 3010,
Australia\\
$^3$Research School of Astronomy \& Astrophysics, Australian National
University, Weston Creek, ACT 2611, Australia}

\begin{document}
\maketitle

\begin{abstract} 
We present the results from a multi-epoch survey of two regions of M33 using the 3.5m WIYN telescope. The inner field is located close to the centre of the galaxy, with the outer region situated about $5.1$ kpc away in the southern spiral arm, allowing us to sample a large metallicity range. We have data for 167 fundamental mode Cepheids in the two regions. The reddening-free Wesenheit magnitude $W_{vi}$ period-luminosity relations were used to establish the distance modulus of each region, with $\mu_{inner} = 24.37 \pm 0.02 \text{ mag}$ and $\mu_{outer} = 24.54 \pm 0.03 \text{ mag}$. The apparent discrepancy between these two results can be explained by the significant metallicity gradient of the galaxy. We determine a value for the metallicity parameter of the Period--Luminosity relation $\gamma = \dfrac{\delta(m-M)_{0}}{\delta\text{log}Z}$ = $-0.29 \pm 0.11 \text{\ mag dex}^{-1}$, consistent with previous measurements. This leads to a metallicity corrected distance modulus to M33 of $\mu_{\gamma} = 24.53 \pm 0.11$ mag.
\end{abstract}


\section{Introduction}
\label{introduction}

Cepheid variables have been used to determine galactic distances for
many decades (e.g. \citet{1969PASP...81..707F},
\citet{2001ApJ...553...47F} and references therein). The relationship
between their period of pulsation and luminosity has been considered
to be well defined at several times during its lifetime. However, the
effect of the Cepheid's chemical composition on the Period-Luminosity
(PL) relation has never been adequately determined, leaving the
relation open to systematic uncertainties that have not been fully
investigated. Proper calibration of this relation is essential not
just to explain how metallicity affects the pulsations of Cepheids,
but to tie down the zero point of the extra-galactic distance scale
and thus properly measure the Hubble constant.

Metallicity and differential absorption will both change stellar colours \citep{1969MNRAS.142..161B}. Theoretical models
(e.g. \citet{2000A&A...359.1059C}) show that the metal content of a
Cepheid will affect the derived distance modulus. If a star of an
unknown metallicity is at an undetermined distance then it is
difficult to untangle these two effects on the stars colours. Ideally
the PL relation would remove this problem as the distance to the star
could be found without calculating the colour of the Cepheid. The
problem arises as the calibration of the PL relation does not take
this effect into account. The distances that were used to calibrate
the zero point of the relation were determined without considering how
the reddening of the Cepheids was affected by the combination of these
two mechanisms, meaning that a systematic offset may have been
introduced into the relation.

There has been some investigation into the effect that metallicity has
on the PL relation, beginning with \citet{1990ApJ...365..186F}. \citet{1997A&A...318L..47B} and
\citet{1997A&A...324..471S} used the Cepheids detected in the EROS
gravitational lensing survey to show that discrepant values of $H_0$
could be reconciled by applying a metallicity correction to Cepheid
distance moduli.  \citet{1998ApJ...498..181K} used the Hubble Space
Telescope to observe Cepheids in M~101, a nearby spiral galaxy with
active star formation regions but low dust extinction
properties. Choosing a galaxy with low extinction means that there
would be less reddening from dust; any reddening of the colour of the
Cepheids could be explained as a metallicity effect. This study found
that $\gamma = -0.24 \pm 0.16$ mag dex$^{-1}$ (see also Freedman et
al., 2001).
However, their conclusions were based on small numbers of Cepheids,
combined with the PL relations from \citet{1999AcA....49..223U}. It is
far from ideal to use such a small sample to quantify an effect that
could mean that the entire extra-galactic distance scale is subject to
a systematic offset. Using a similar method but with a larger sample
of Cepheids, \citet{2006ApJ...652.1133M} found $\gamma = -0.29 \pm
0.09$ mag dex$^{-1}$.  Using a large sample of nearly 700 Cepheids in
17 galaxies,  \citet{1997ApJ...491...13K} had found a result
compatible with these other works.

Most works use a sample of Cepheids covering a range in metallicity
and enforce consistency between distance moduli; this is what
\citet{1998ApJ...498..181K} and \citet{2006ApJ...652.1133M} did for
instance.  An alternative approach used e.g. by
\citet{2004ApJ...608...42S} is to assume that the distances to several
galaxies are known a priori, from the tip of the red giant branch in
their case, and find the correction needed to bring Cepheid distances
to these galaxies in agreement with that other standard candle.

There are two possible solutions to the problem of the
metallicity-extinction degeneracy. The method followed by
\citet{1998ApJ...498..181K} is one possibility, the elimination of
extinction from the equation. Theoretically this method should work,
but in practice it is not so simple. \citet{1998ApJ...498..181K} chose a 
galaxy with low extinction, not zero extinction. This implies that
the extinction is not entirely eliminated, leaving the degeneracy that
we are trying to overcome. However, this degeneracy could be removed
by using the reddening-free Wesenheit magnitude $W_{vi}$ PL relation
to calculate the distance. By eliminating extinction in this way, any
discrepancies in the apparent distance modulus could be explained by
the metallicity effect.

The work presented here is based on a multi-epoch survey of M33
conducted with the WIYN 3.5 m telescope. M33 is a well studied galaxy,
with observations of Cepheids performed by Hubble as early as 1926
\citep{1926ApJ....63..236H}. As it is nearby, many studies have
determined the distance of M33 using different and independent
methods, such as eclipsing binaries \citep{2006ApJ...652..313B}, the
tip of the red giant branch (TRGB) \citep{2004AJ....128..237B,2004A&A...423..925G,2004MNRAS.350..243M}, masers
\citep{2005Sci...307.1440B} and Cepheids themselves \citep{1991ApJ...372..455F}. Although some of these methods may not
provide the same accuracy that a Cepheid distance does, they allow us
to have an independent starting point for our analysis of the PL
relation of M33. One of the reasons that M33 was chosen for this study
is that it is well studied. Its inclination angle is 53$^\circ$, and so any effect 
on distance modulus ($\mu$) from the orientation will be extremely small. The effect on $\mu$
due to the orientation of the galaxy was found to be very small (less than 0.01 mag). The
metallicity gradient across the galaxy is well defined, so the
chemical composition of the Cepheids is relatively easy to determine
just from their position in the galaxy. These properties inherent to
M33 make it the ideal target for a study of this nature.


The paper is organised as follows. Section~\ref{observations}
describes our observations. Section~\ref{data-reduction} describes the
reduction process, including the development of the photometry
pipeline and the final calibration and
astrometry. Section~\ref{results} contains the final catalogues and
assesses the accuracy of the photometry, with the final PL relations
and distances in section~\ref{pl-relation}. Finally, we examine the
effect of metallicity on our results and derive a value for the
metallicity correction parameter $\gamma$.

\section{Observations}
\label{observations}

The data set consists of 25 epochs of $BVI_{C}$ images of four fields
of M33 obtained between 1998 and 2001 on the 3.5 m WIYN telescope. The
first 20 epochs used the WIYN s2kb Imager, a single thinned Tek/STIS
2048x2048 CCD with 21 $\mu$m pixels. For the final 5 epochs the new
Mini-Mosaic (MiMo) camera was used. The Mini-Mosaic consists of two
thinned SITe 2048x4096 CCDs with 15 $\mu$m pixels. A single exposure
covered $6.8 \times 6.8$ arcmin for the Imager, and $9.6 \times 9.6$
arcmin for the Mini-Mosaic, with pixel scales of 0.195 and 0.141 arcsec pixel$^{-1}$ 
respectively. 

\begin{figure}
\includegraphics[width=85mm]{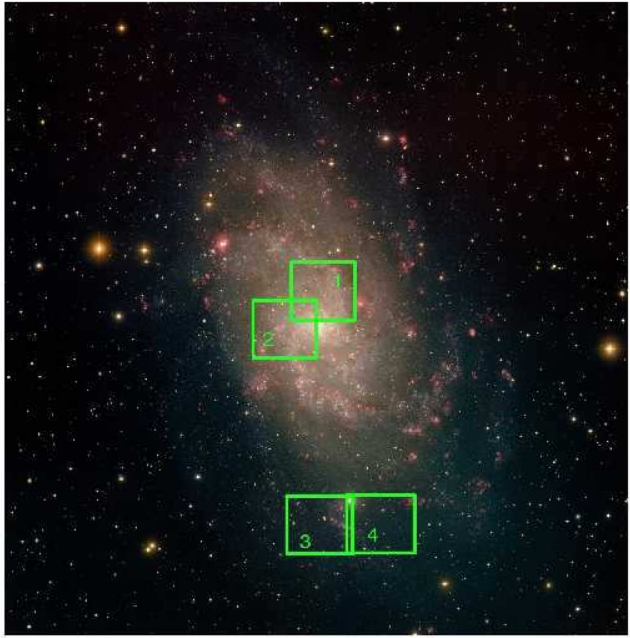}
 \caption{The two regions observed for this survey\protect\footnotemark[1]. The inner region contains fields 1 and 2; the outer region contains fields 3 and 4. The fields were chosen such that they would cover the whole metallicity range of the galaxy. Orientation: North up, East right.}
 \label{fields-figure}
\end{figure}
\footnotetext[1]{Original image copyright: National Optical Astronomy Observatory/Association of Universities for Research in Astronomy/National Science Foundation}

Four fields were observed for the survey, two in the centre of M33, and two in the outer, southern part of the galaxy. The locations of the fields are shown in Figure~\ref{fields-figure}. The J2000.0 coordinates of the fields are shown in Table~\ref{field-table}. All the fields were observed in the $B$, $V$ and $I_{C}$ filters with both cameras. The position of the fields were chosen such that the metal-rich population at the centre of the galaxy could be compared with the more metal-poor population in the outer region. Cepheids tend to be concentrated in the spiral arms, so the outer fields were located in the southern part of the galaxy to coincide with the clearest arm. The positions of the fields and the addition of the Mini-Mosaic camera meant that some objects were observed in either both inner or both outer fields, giving up to 38 observations for objects in the overlap regions.

\begin{table}
\caption{J2000.0 coordinates of the four fields. M33-1 and M33-2 are situated in the centre of the galaxy. M33-3 and M33-4 are in the southern spiral arm. Deprojected angular and linear distances of each field to the centre of the galaxy are given, assuming M33 is at a distance of 840 kpc, has an inclination angle of $53^{\circ}$ and the major axis has a position angle of $22^{\circ}$.}
\begin{center}
\begin{tabular}{l c c c c}
\hline \hline
Field & $\alpha$ & $\delta$ & Angular & Linear \\ \hline
Imager & & & & \\ \hline
M33-1 & $1^{h} 33^{m} 49.12^{s}$ & $30^{\circ} 42' 56.48''$  & $217.64''$ & $0.886$ kpc \\
M33-2 & $1^{h} 34^{m} 06.13^{s}$ & $30^{\circ} 38' 52.50''$ & $217.10''$ & $0.884$ kpc\\
M33-3 & $1^{h} 33^{m} 48.99^{s}$ & $30^{\circ} 20' 24.58''$ & $1242.13''$ & $5.059$ kpc\\
M33-4 & $1^{h} 33^{m} 25.63^{s}$ & $30^{\circ} 20' 25.25''$  & $1290.30''$ & $5.255$ kpc\\
\hline
Mi-Mo & & & & \\
\hline
M33-1 & $1^{h} 33^{m} 51.77^{s}$ & $30^{\circ} 45' 16.56''$ & $367.50''$ & $1.497$ kpc\\
M33-2 & $1^{h} 34^{m} 16.75^{s}$ & $30^{\circ} 36' 35.64''$ & $409.13''$ & $1.666$ kpc\\
M33-3 & $1^{h} 34^{m} 00.16^{s}$ & $30^{\circ} 21' 46.08''$ & $1161.16''$ & $4.729$ kpc\\
M33-4 & $1^{h} 33^{m} 21.13^{s}$ & $30^{\circ} 21' 52.20''$ & $1220.28''$ & $4.970$ kpc\\
\hline

\end{tabular}
\end{center}
\label{field-table}
\end{table}

\begin{table}
\caption{Log of Observations. Table~\ref{obs-table} is available in full electronically. Cameras: IM = S2KB Imager, MiMo = Mini-Mosaic}
\begin{center}
\begin{tabular}{l c c c c c}
\hline \hline
UT Date & MHJD & Field & Filter & Cam. & Seeing \\ \hline
1998 Aug 19 & 51044.427 & M33-1 & $B$ & IM & 0.53'' \\ 
1998 Aug 19 & 51044.433 & M33-1 & $I_{C}$ & IM & 0.53'' \\ \hline
\end{tabular}
\end{center}
\label{obs-table}
\end{table}

\section{Data Reduction}
\label{data-reduction}
Preliminary reduction of the data were performed using the routines in the {\sc iraf}\footnote{{\sc iraf} is distributed by the National Optical Astronomy Observatories, which are operated by the Association of Universities for Research in Astronomy, Inc., under cooperative agreement with the National Science Foundation.} {\sc ccdproc} package. This consisted of overscan correction, bias subtraction and flat fielding. The photometry was performed using the standalone {\sc daophot} \citep{1987PASP...99..191S} and {\sc allframe} \citep{1994PASP..106..250S} packages kindly provided by P. Stetson. 

\subsection{Photometry Pipeline}
\label{pipeline}
The data set used in this project is fairly large, containing over 300 images of the four fields. To process such a large data set a completely automatic photometry pipeline was developed, covering all the steps from initial star selection to the final PSF photometry.

Initial star lists were compiled using the {\sc find} routine in {\sc daophot} with a 3$\sigma$ detection limit. Aperture photometry was performed on each list to get rough magnitudes using an 11 pixel aperture and sky values from the annulus from 19 to 23 pixels. 


The creation of the point spread function (PSF) model is the most important part of the process. The fields are extremely crowded; the inner field has an average density of 890 stars arcmin$^{-2}$, or approximately 10 pixels between each centroid position. Of course these calculations do not take into account the clumpiness of the distribution, the galactic centre has a much higher stellar density than further out. As the fields are so crowded it is imperative that the PSF model is as accurate a representation of the stellar profile as possible, but it also means that it becomes a more taxing problem. The stars chosen to make the model should be the brightest, least crowded available, and with such heavy crowding these are very rare objects.

The problem was overcome using a combination of the {\sc daophot} {\sc pick} and {\sc psf} routines along with a series of scripts to analyse the resulting fits. The initial model was created by allowing {\sc pick} to choose the 200 best stars from the aperture photometry file then using {\sc psf} to make six model profiles using the different Gaussian, Moffat and Lorentz functions available, allowing the model to vary linearly across the frames. The residuals of the fit were analysed and any stars flagged as saturated or with fitting residuals more than twice the root-mean-squared (RMS) residual value were removed. All the stars except the PSF stars were then removed from the images and the PSF was remade with the new star list from the cleaned image. The process of analysing the residuals, removing neighbour stars and remaking the PSF was repeated a further five times and resulted in a clean, well fitting model. Allowing the model to vary quadratically and repeating the cleaning process more times were both tested but did not significantly improve the results.

Profile fitting photometry was performed on each frame separately using the {\sc allstar} routine, which produced an image with all the fitted stellar profiles subtracted. A fitting radius of 0.78 arcsec was used, corresponding to the typical seeing. This new image was run through the star finding process again to detect any stars that were too faint or crowded to be seen in the first pass. This new star list was appended to the original and run through {\sc allstar} again.

Once reasonable PSF magnitudes had been obtained for each frame, the
images were combined using {\sc montage2} (Stetson, personal
communication) using the best 50 frames to produce two master images,
one for the inner and another for the outer fields. These deep,
medianed images were then run through the pipeline described above to
produce master star lists for the two regions. The master lists
contained over 200,000 stars in the inner region and around 100,000
for the outer region. The lists produced from these frames were used
rather than the original lists for two reasons. Firstly, a medianed
image made from many frames has a much better signal to noise ratio (S/N)
than a single frame. Secondly, using identical star lists for
each frame and each filter is necessary when using the {\sc allframe}
package. {\sc allframe} was used to calculate the final PSF magnitudes from
`fixed position' coordinate lists.

The final step of the instrumental photometry was to calculate the aperture corrections and set all the photometry to the same zero point. Landolt standards were observed on 1999 October 3rd so this night was chosen to be the reference night. Aperture corrections were calculated for each frame taken on the reference night (4 fields in 3 filters) using {\sc daogrow} \citep{1990PASP..102..932S}. Using the {\sc allframe} photometry files as input, all the stars in each frame except the stars used to make the PSF model were subtracted. The PSF stars are ideal for calculating the aperture correction as they are by their nature bright and relatively isolated. The stars remaining in the frame were then measured in nine apertures with radii between 4 and 20 pixels, with the sky measured from an annulus between 25 and 30 pixels. The output photometry was then put into {\sc daogrow} \citep{1990PASP..102..932S} to calculate the growth curves and derive the total magnitudes. The difference between the total magnitude and the PSF magnitude for each star was calculated, and any bad outliers were removed. These stars could have been affected by nearby stars that had been missed, or even nearby PSF stars in the large aperture. The best value of the correction was found using a least-squares fit. The corrections were then subtracted from the PSF magnitude to put the {\sc allframe} photometry on the same scale as the aperture magnitudes. The aperture corrections for the reference night had an average value of 0.035 mag. 

To get correct photometry for every night the zero-point offset between each frame and its corresponding reference frame was calculated. As the fields overlap, field 1 was chosen as the reference for the inner region and field 3 for the outer region. This ensured that stars that were observed in different fields on the same night were on the same zero point. The zero-point offset of each frame was found by taking the average difference between each magnitude on the reference frame and the frame in question, typically using 1000--2000 stars. The star furthest away from the average was removed and the average recalculated until the dispersion of the differences was less than 0.01 mag. This was found to be a very robust method, distinguishing zero-point differences down to one hundredth of a magnitude.

\subsection{Photometric Calibration and Astrometry}
\label{calibration}

Observations of standard fields \citep{1992AJ....104..340L} were taken
on 1999 October 3rd with the Imager CCD. Three standard fields were
observed repeatedly, with a total of 17 different standard
stars. Aperture photometry was performed on the standards with the
same settings as the aperture correction, that is a 20 pixel aperture
and a 25 to 30 pixel sky annulus. The {\sc photcal} routines in {\sc
iraf} were used to construct standard solutions of the form:
\begin{equation} 
m_{inst} = m_{cal} + a + b \times X_{M} + c \times Colour + d \times UT 
\label{cal-1}
\end{equation}
where $m_{inst}$ is the instrumental magnitude, $m_{cal}$ is the calibrated magnitude, $X_{M}$ is the airmass and $Colour$ corresponds to $(B-V)$ for the $B$ and $V$ frames and $(V-I_{C})$ for $I_{C}$. The coefficients for each transformation equation are shown in Table~\ref{Phot-sol-tab}. The solution was applied to all Imager frames using {\sc invertfit}. The $UT$ term was included in the solution as the extinction changed very slowly over the reference night. Although the coefficient of this term in the solution is much smaller than the others we believe that it is still important to include it. The aim of the process is to get the most accurate photometry possible, so if an effect is present and can be measured on a scale comparable to the accuracy we are trying to achieve then it should be considered.  


\begin{table}
\caption{Photometric solution of 1999 October 3rd for the Imager CCD}
\begin{center}
\begin{tabular}{l c c c c c }
\hline \hline
Filter & a & b & c & d & RMS\\ \hline
$B$ & 24.1552 & 0.2760 & --0.0682 & --0.0062 & 0.0098\\ 
$V$ & 24.3051 & 0.2006 & --0.0123 & --0.0062 & 0.0107\\
$I_{C}$ & 23.7141 & 0.0903 & --0.0545 & --0.0026 & 0.0187\\ \hline
\end{tabular}
\label{Phot-sol-tab} 
\end{center}
\end{table}

This calibration is only appropriate for the frames taken with the
Imager CCD. Once applied to the Imager frames, the stars on the
reference frames could be used as local standards to derive a
photometric solution for the two Mini-Mosaic CCDs. The reference night
for the Mini-Mosaic was chosen to be 2000 October 3rd as this had the
best image quality for all fields. Stars that appeared in all frames
of field 3 on both the photometric reference night and on the Mosaic
reference night with $\sigma_{V} \leq 0.02$ mag were selected as the
local standards -- around 200 stars for each chip. The {\sc photcal}
routines were used to find the solution of the transformation
equations from the Mosaic instrumental magnitudes to the Imager
Landolt calibration. For this transformation the time dependence and
airmass term are absorbed into the zero-point
coefficient of this solution and a quadratic colour
term was added, resulting in transformation equations of the form
\begin{equation}
m_{MiMo} = m_{Im} + a + b \times Colour + c \times (Colour)^{2}
\label{cal-2}
\end{equation}
where $m_{MiMo}$ refers to the magnitude on the Mini-Mosaic
instrumental scale and $m_{Im}$ refers to the calibrated Imager scale.
The quadratic colour term of the Mini-Mosaic camera is a feature that
has been described before by \citet{2005PASP..117..563S}.  The
coefficients of the equations are shown in Table~\ref{Mos-sol-tab}.


\begin{table}
\caption{Photometric solution of 2000 October 3rd for the Mini-Mosaic Camera}
\begin{center}
\begin{tabular}{l c c c c }
\hline \hline
Filter & a & b & c & RMS\\ \hline
\textit{Chip 1}  \\ \hline
$B$ & 23.9118 & --0.0935 & 0.0172 & 0.0494 \\
$V$ & 24.1484 & --0.0235 & 0.0079 & 0.0359 \\
$I_{C}$ & 23.6443 & --0.0374 & -0.0083 & 0.0472 \\
\hline
\textit{Chip 2} \\ \hline
$B$ & 23.9120 & -0.0587 & --0.0143 & 0.0486 \\
$V$ & 24.1465 & 0.0128 & --0.0245 & 0.0302 \\
$I_{C}$ & 23.6373 & --0.0610 & 0.0042 & 0.0455 \\
\hline
\end{tabular}
\label{Mos-sol-tab} 
\end{center}
\end{table}

The stacked frames created for making the {\sc allframe} master lists were also used to perform the astrometry. For the outer field the USNO-B catalogue \citep{2003AJ....125..984M} were used, resulting in a solution with RMS values of $0.4$''. The USNO-B catalogue was not appropriate for the inner field as it covers the central part of the galaxy, which is extremely crowded. For this region the astrometric solution was found by matching by eye to the catalogues published in \citet{2006AJ....131.2478M}. They also use USNO-B catalogue for this region but have many fewer stars. This process gave a much better solution, with an RMS below $0.07$''.

\section{Results}
\label{results}

\subsection{The Catalogues}
\label{catalogues}

\begin{figure*}
$\begin{array}{c@{\hspace{10mm}}c}
\includegraphics[width=84mm]{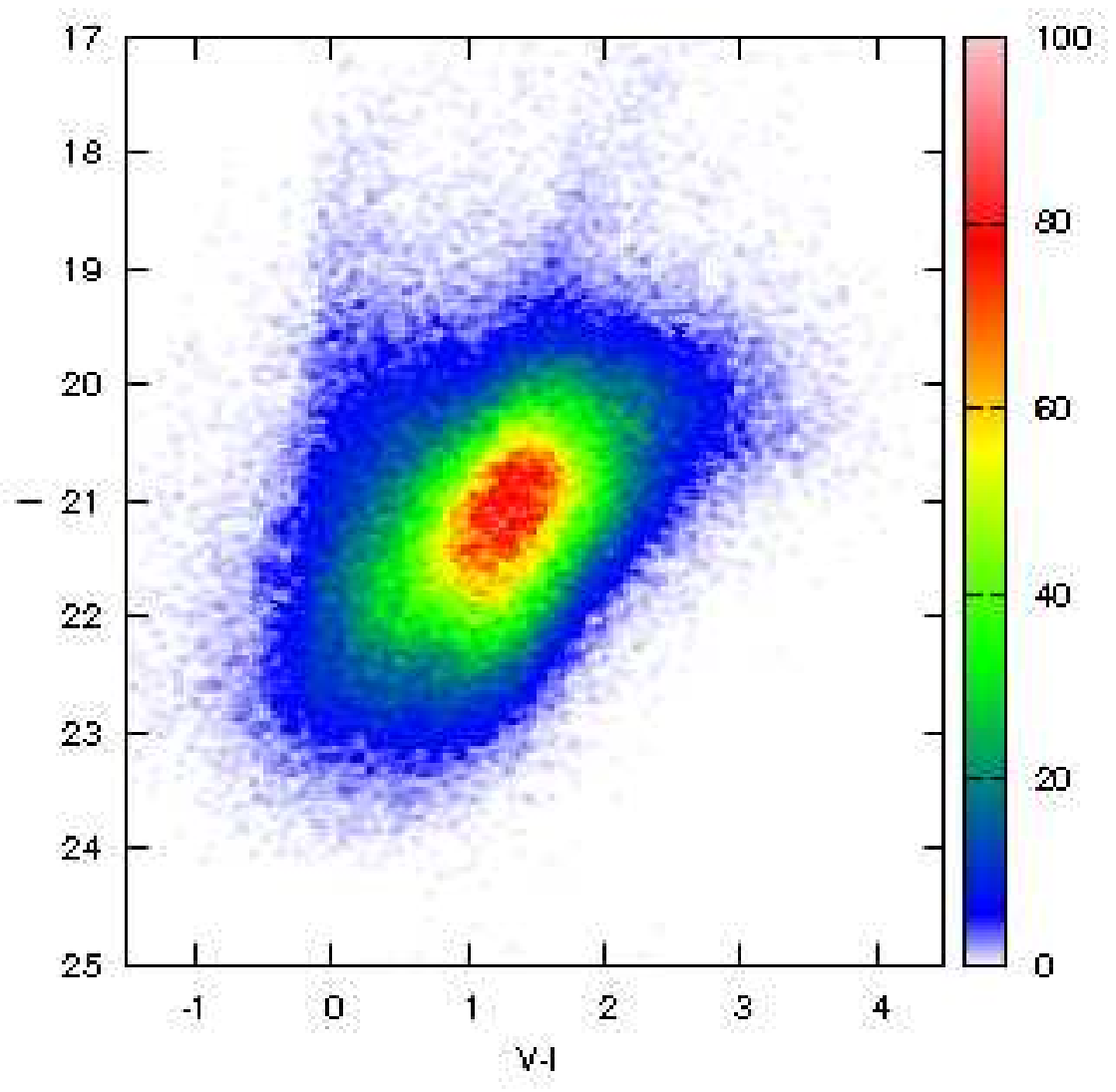} &
\includegraphics[width=84mm]{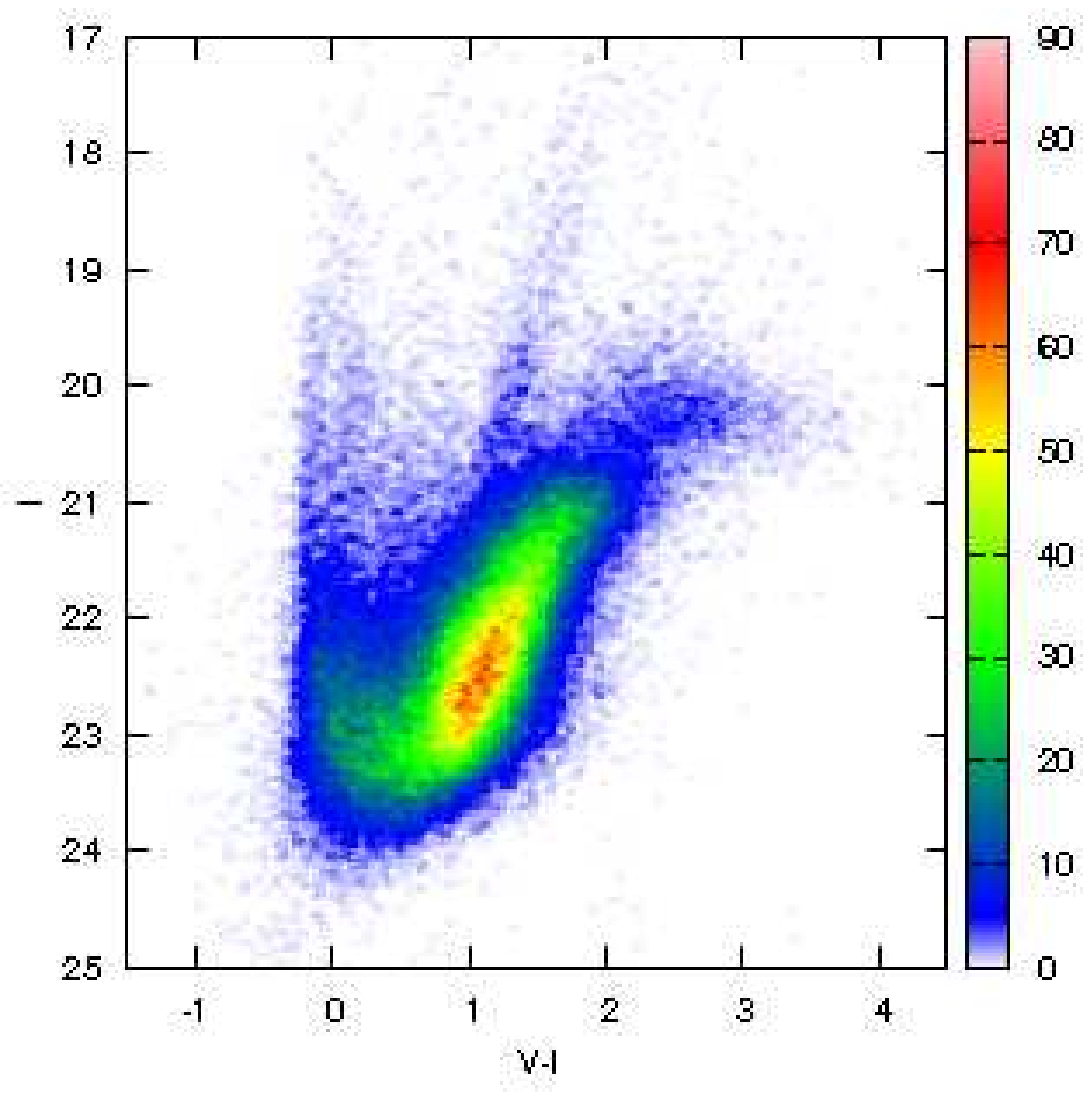} \\
\end{array}$
\caption{Hess diagrams for the inner (left) and outer (right) fields, with 0.025 mag
  bins. The inner field does not go as deep as the outer field as the
  degree of crowding is much higher. The effects of crowding and
  differential reddening are obvious in the inner field but are much less so in
  the outer field}
 \label{cmds-figure}
\end{figure*}

Two catalogues are available from the WIYN M33 Survey. The first
contains weighted average $B$, $V$ and $I_{C}$ magnitudes for each
star. A sample of this catalogue is shown in Table~\ref{Cat-1}\footnote{The full catalogues are available on request from the authors.}. The
second contains all the individual measurements for all the variables
cross-identified with the Deep CFHT Photometric Survey of M33
\citep{2006MNRAS.371.1405H}.  The catalogues are available in full
electronically. They cover the magnitude range between $13.8
\leq V \leq 26.7$, with 75465 and 146820 in the final outer and inner
field lists respectively. The magnitude range of the inner field is
limited slightly by the degree of crowding, and only contains objects
down to $V=25.7$. Colour-magnitude diagrams for both fields are shown
in Figure~\ref{cmds-figure}.

\begin{table*}
 \centering
\begin{minipage}{160mm}
 \caption{Sample of the weighted averages catalogue. The uncertainties are from the weighted average calculations derived from the {\sc allframe} results. Where no observations were made the magnitude and error are represented by 99.999 and 9.999 respectively.}
  \begin{tabular}{l c c c c c c c c c c c }
\hline \hline
ID & $\alpha$ hh:mm:ss & $\delta$ dd:mm:ss & $B$ & $\sigma_{B}$ & $V$ & $\sigma_{V}$ & $I_{C}$ & $\sigma_{I}$ & $N_{B}$ & $N_{V}$ & $N_{I}$ \\ \hline
OUT-14501 & 1:33:10.21 & 30:17:22.7 & 24.946 & 0.423 & 23.733 & 0.161 & 99.999 & 9.999 & 1 & 1 & 0 \\
OUT-14510 & 1:33:10.25 & 30:17:05.8 & 22.484 & 0.155 & 22.269 & 0.096 & 99.999 & 9.999 & 2 & 2 & 0 \\
OUT-14523 & 1:33:10.26 & 30:17:07.2 & 22.525 & 0.092 & 22.872 & 0.075 & 99.999 & 9.999 & 2 & 2 & 0 \\
OUT-14527 & 1:33:10.19 & 30:17:50.4 & 23.608 & 0.312 & 24.306 & 0.248 & 99.999 & 9.999 & 1 & 1 & 0 \\
OUT-14532 & 1:33:10.26 & 30:17:15.2 & 23.475 & 0.109 & 22.943 & 0.067 & 99.999 & 9.999 & 2 & 2 & 0 \\
OUT-14539 & 1:33:10.27 & 30:17:17.2 & 23.474 & 0.125 & 23.576 & 0.098 & 23.232 & 0.335 & 2 & 2 & 1 \\
OUT-14546 & 1:33:10.27 & 30:17:23.2 & 24.193 & 0.184 & 23.556 & 0.092 & 23.002 & 0.219 & 2 & 2 & 1 \\
OUT-14551 & 1:33:10.14 & 30:18:33.5 & 24.732 & 0.373 & 23.040 & 0.129 & 99.999 & 9.999 & 1 & 1 & 0 \\
OUT-14554 & 1:33:10.17 & 30:18:20.5 & 22.238 & 0.096 & 22.313 & 0.075 & 99.999 & 9.999 & 1 & 1 & 0 \\
OUT-14555 & 1:33:10.20 & 30:18:06.2 & 24.940 & 0.410 & 23.103 & 0.084 & 99.999 & 9.999 & 1 & 1 & 0 \\ \hline
\end{tabular}
\label{Cat-1}
\end{minipage}
\end{table*}

\begin{figure}
\includegraphics[width=84mm]{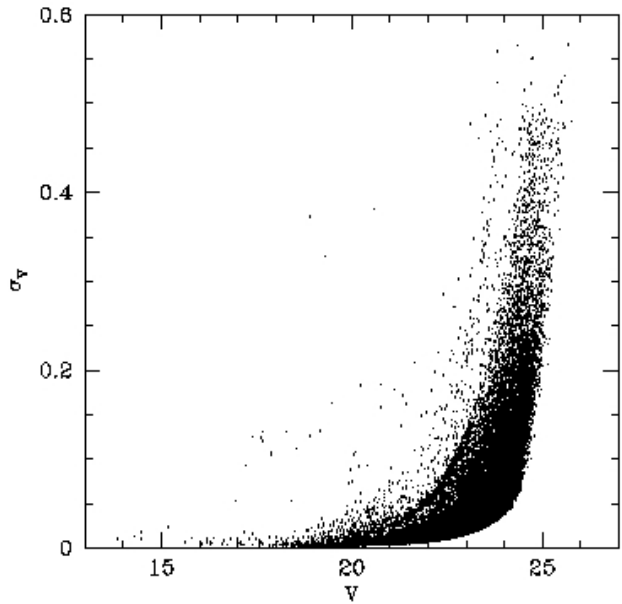}
 \caption{Magnitude - $\sigma$ relation for $V$ in the outer field. The second series with slightly higher uncertainties for a given magnitude is due to the Mini-Mosaic magnitudes dominating these objects, as these stars have few measurements.}
 \label{error-figure}
\end{figure}

The uncertainties given in the average photometry catalogue are the
errors on the weighted average magnitudes, calculated from the {\sc allframe}
 results. Figure~\ref{error-figure} shows the relation
between magnitude and uncertainty for the $V$ band in the outer
field. A second series becomes visible around $m_V=20$ mag.  These
objects have fewer measurements, a maximum of five, as they appear
only on the Mini-Mosaic frames (which has a large field of view than
the Imager).
However, a comparison between magnitudes calculated from the Imager
and the Mini-Mosaic shows the photometry is compatible. Neglecting the
second series we obtain 1\% photometry for objects brighter than
$m_V=22$, and 10\% down to $m_V=24$. The full analysis is shown in
Table~\ref{errors-table}. These results are compatible with
theoretical estimates from signal to noise calculations.

\begin{table}
 \caption{Typical magnitudes and uncertainties from the weighted averages catalogue.}
\begin{center}
 \begin{tabular}{l c c c} 
  \hline \hline
Field & Filter & $\sigma$ = 1\% & $\sigma$ = 10\% \\ \hline
Inner & $B$ & 21.125 & 24.129 \\
 & $V$ & 21.693 & 23.854 \\
 & $I_{C}$ & 19.896 & 22.597 \\
Outer & $B$ & 21.336 & 24.404 \\
 & $V$ & 22.136 & 24.195 \\
 & $I_{C}$ & 20.840 & 22.643 \\
\hline
 \end{tabular}
\end{center}
\label{errors-table}
\end{table}

\subsection{Comparison with others}
\label{comparison}

\begin{figure*}
 \centering
\begin{minipage}{180mm}
$\begin{array}{c@{\hspace{10mm}}c}
\includegraphics[width=84mm]{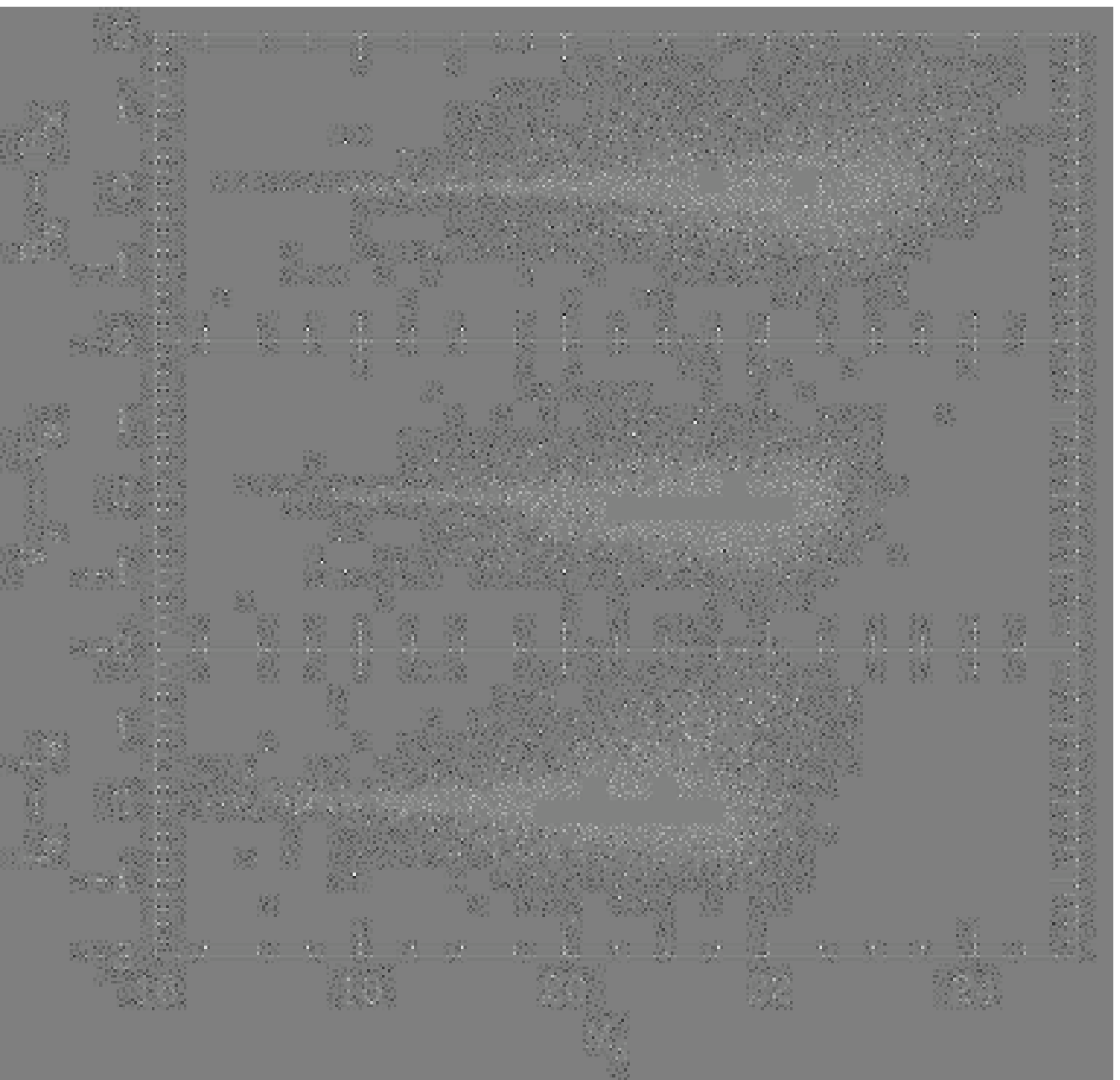} &
\includegraphics[width=84mm]{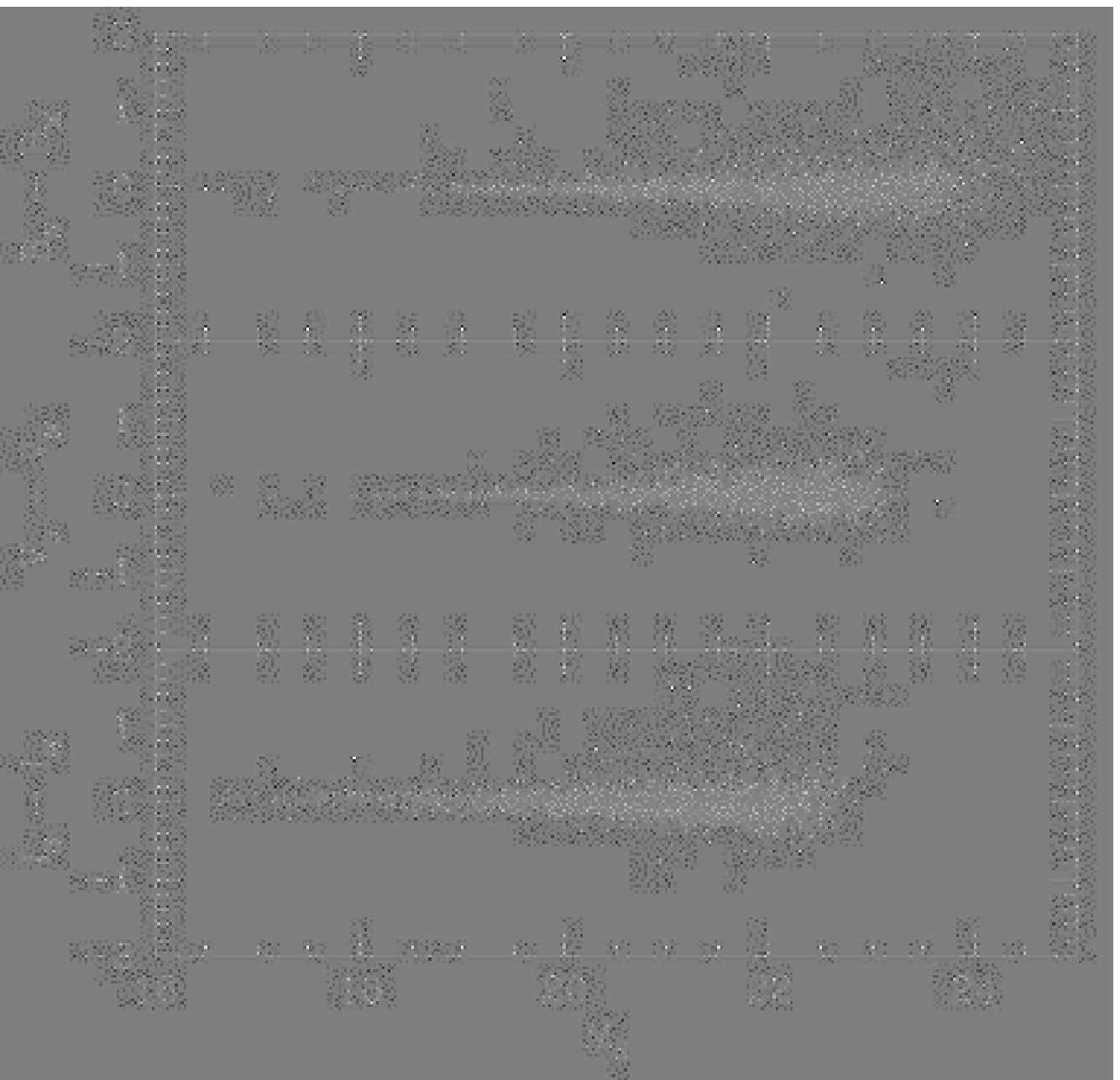} \\
\end{array}$
\caption{Comparison between this work and the catalogues of \citet{2006AJ....131.2478M} for the inner field (left) and the outer field (right). This work is considered to be the reference and is denoted by $M_{S}$, \citet{2006AJ....131.2478M} by $M_{M}$ and $\Delta mag = M_{S} - M_{M}$. The average number of stars compared is 9700 for the inner field and 5100 for the outer field, corresponding to roughly 5\% of our catalogues. All matched stars with $\sigma < 0.5$ mag that were outside the central 4 arcmin were included. The secondary $M_{S}-M_{M} < 0$ trends that are visible in the left hand panel are believed to be due to the severe crowding of the region, as the matching algorithm used will tend to match to slightly brighter stars in highly crowded regions.}
\label{comparison-fig}
\end{minipage}
\end{figure*}

To check the accuracy of our results the average magnitudes were compared with those obtained by \citet{2006AJ....131.2478M}. Their survey covers a larger area than the survey presented here but does not go as deep. The comparison for each filter in both fields is shown in Figure~\ref{comparison-fig}.  The photometry from this work is considered as the reference, $\Delta mag = m_{Scowcroft} -m_{Massey}$. These  plots show that the photometry is sound.  Most objects have a difference near zero as expected.  For the inner field in particular, there seems to be  a secondary sequence near $\Delta mag \approx -0.8$.  We attribute this to misidentifications due to the high level of crowding. By excluding the central 4 arcmin of the galaxy from the comparison the effect was significantly reduced. 

The average differences were found using a least squares fit to the brightest three magnitudes in the comparison and are shown in Table~\ref{diffs-tab}. Only the brightest stars in the comparison were chosen as the two catalogues are optimised for different magnitude ranges; were we to consider the whole range we would be basing the fit on the very faintest stars in \citet{2006AJ....131.2478M}. However, $\Delta mag$  gradients were checked for, and the differences for each field-filter combination were found to be constant over the whole range of magnitudes. We find a larger average $\Delta mag$ for $I_{C}$ which is believed to be because the $I_{C}$ frames are more crowded than $B$ and $V$, particularly in the inner field.

\begin{table}
 \begin{center}
\caption{Differences between weighted average magnitudes in this
  catalogue and \citet{2006AJ....131.2478M}. $\Delta mag$ was found
  from a least squares fit to the brightest end of the comparison, the
  range is given in column 4. $\sigma_{\Delta mag}$ is the uncertainty on $\Delta mag$. Outliers with $\Delta mag > \pm 0.5$
  were rejected, as they are most likely misidentifications due to the
  crowding in the inner field. \label{diffs-tab}}
  \begin{tabular}{l c c c c}
   \hline \hline
Field & Filter & $\Delta mag$ & $\sigma_{\Delta mag}$ & Range\\ \hline
Inner & $B$ & 0.008 & 0.005 & 16-19 \\
 & $V$ & --0.009 & 0.008 & 15-18 \\
 & $I_{C}$ & 0.031 & 0.008 & 15-18 \\
Outer & $B$ & --0.013 & 0.008 & 16-19\\
 & $V$ & --0.014 & 0.005 & 16-19\\
 & $I_{C}$ & 0.021 & 0.005 & 16-19\\
\hline
  \end{tabular}
 \end{center}
\end{table}

\section{Cepheid Period-Luminosity Relation}
\label{pl-relation}
The aim of the survey was to detect Cepheids in the two fields and assess the effect that metallicity has on the PL relation. The positions of the two fields were chosen such that the inner field covered the metal-rich centre, which has an [Fe/H] value slightly larger than the Milky Way (MW), and the outer field covered the southern spiral arm where the metallicity drops to slightly below that of the Large Magellanic Cloud (LMC). 

\subsection{Cepheid Selection}
\label{selection}
Cepheids were identified by cross referencing the catalogue with the fundamental mode Cepheid catalogue produced as part of the deep CFHT survey of M33 \citep{2006MNRAS.371.1405H}. The fundamental mode Cepheids were identified using Fourier parameters \citep{1981ApJ...248..291S}. First overtone Cepheids were excluded from the analysis as they are known to sit slightly higher in the $\log P - Mag$ plane than fundamental Cepheids and would bias any zero-point fitting towards brighter magnitudes. 

To ensure an accurate PL relation the average magnitudes for each Cepheid were recalculated by fitting Fourier series to the light curves, using the form shown in Equation \ref{fourier-eq}, where $m(\phi)$ is the magnitude at phase $\phi$ and $a$ becomes the phase averaged magnitude. 
\begin{equation}
 m(\phi) = a + \displaystyle\sum_{i=1}^3 b_{i} cos (2 \pi i \phi + c_{i})
\label{fourier-eq}
\end{equation}
The light curve of each Cepheid was examined visually to remove any misidentifications or ones with incorrect periods. The amplitude of each Cepheid was calculated and Cepheids with $A_{B} > A_{V} > A_{I_{C}}$ were included in the final sample, similar to the selection criteria of \citet{2002AJ....123..789P}. Amplitude ratio selection is an important step as large numbers of classical Cepheids are known to exist in binary systems \citep{2003ASPC..298..237S}. Very close companions would not be identified as separate stars in the initial photometry. They can however be identified by their characteristic effect on the Cepheid light curve. A Cepheid with an unidentified companion will still appear variable, but the light curve would not reach the minimum value expected. The effect will be seen to different extents in the different photometric bands depending on the colour of the companion. This explanation can also be extended to blending; if a Cepheid is within a background of unidentified faint stars these objects will affect the shape of the light curve in the same way. By requiring that the amplitudes of the Cepheids follow the criteria set out above we can remove these effects from the sample. The visual inspection and amplitude selection left 91 and 28 Cepheids in the inner and outer field samples respectively. 

\subsection{Metallicity effects}
\label{metallicity}

Many attempts have been made to assess the extent of the metallicity effect, with some analysing the effect of $Z$ on $\mu$ -- and hence the PL zero-point -- such as \citet{2008ApJ...684..102B} and \citet{2004ApJ...608...42S}, whilst others focus on the effect that $Z$ has on the slope of the PL relation (e.g. \citet{1999A&A...344..551A}). In this work we analyse the effect of $Z$ on the derived distance modulus, which is quantified by the parameter $\gamma$:
\begin{equation}
 \gamma = \dfrac{\delta(m-M)_0}{\delta \log Z}
\label{gamma}
\end{equation}
where $\delta (m-M)_{0}$ is the difference between distance modulus corrected for the metallicity effect and without correction, in the sense corrected -- apparent, and $\delta \log Z$ is the difference between a reference metallicity and the region being studied \citep{1998ApJ...498..181K}. Recent measurements of $\gamma$ have produced negative values as large as $\gamma = -0.4$ mag dex$^{-1}$, with the average being around $\gamma = -0.25$ mag dex$^{-1}$. Following the methodology of \citet{1998ApJ...498..181K} and \citet{2004ApJ...608...42S} the value of $\gamma$ can be calculated using the two distance moduli, $\mu_{inner}$ and $\mu_{outer}$, and the [O/H] gradient of M33. We adopt the recent [O/H] gradient from \citet{2007A&A...470..865M}:
\begin{equation}
 \dfrac{d[O/H]}{dR} = -0.19 \pm 0.08 \text{ dex kpc}^{-1} \text{ (R} < \text{3 kpc)}
\label{inner-z}
\end{equation}
\begin{equation}
 \dfrac{d[O/H]}{dR} = -0.038 \pm 0.015 \text{ dex kpc}^{-1} \text{ (R} \geq \text{3 kpc).} 
\label{outer-z}
\end{equation}
The deprojected radial distances of the centres of the two fields are 0.52 kpc and 5.11 kpc, leading to $12 +
\log\text{(O/H)} = 8.852$ and $8.296$ for the inner and outer fields respectively. This gives a value of $\Delta [O/H] = -0.556$. To realise the full extent of the correction, we must initially treat the fields as if they had the same metallicity. This is done by using the same model PL relations for both regions.

PL relations were fitted to the Cepheids in both fields in $B,V,I_{C}$ and the reddening-free Wesenheit magnitude $W_{vi}$ as defined by \citet{2007A&A...476...73F} (F07), assuming the $R_{V}=3.23$ reddening law described by \citet{1989ApJ...345..245C}:
\begin{equation}
 W_{vi} = V - 2.55 (V - I_{C})\text{.}
\label{wvi}
\end{equation}

The PL relations take the form $M = a \log P + b$. The values for the slopes, $a$, were taken directly from the F07 LMC relations, whilst the intercept, $b$, was derived from an iterative weighted least-squares fit for each relation. The values of the $W_{vi}$ zero-points are shown in Table~\ref{different-pls}. 

The zero-points of the PL relations for $W_{vi}$ can be used to directly infer the distance modulus of M33, without the need to correct for extinction. The distance modulus can be found from the apparent $W_{vi}$ magnitude of a Cepheid with a period of one day, such that $\mu = b - M_{0}$, where b is the zero-point of the PL relation and $M_{0}$ is the absolute magnitude of a one day Cepheid, which is taken directly from the LMC PL relations from F07.

\begin{figure*}
 \centering
\begin{minipage}{180mm}
$\begin{array}{c@{\hspace{10mm}}c}
\includegraphics[width=84mm]{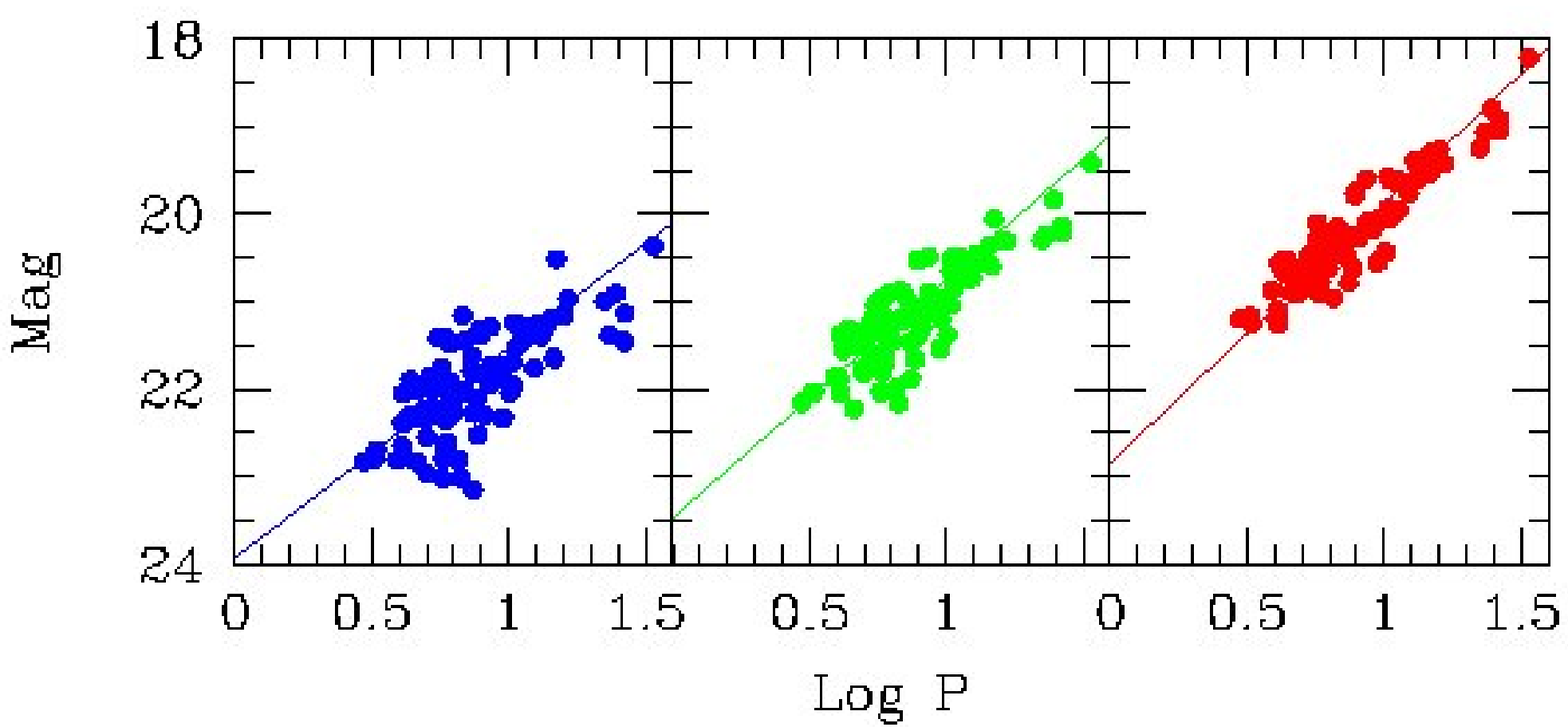} &
\includegraphics[width=84mm]{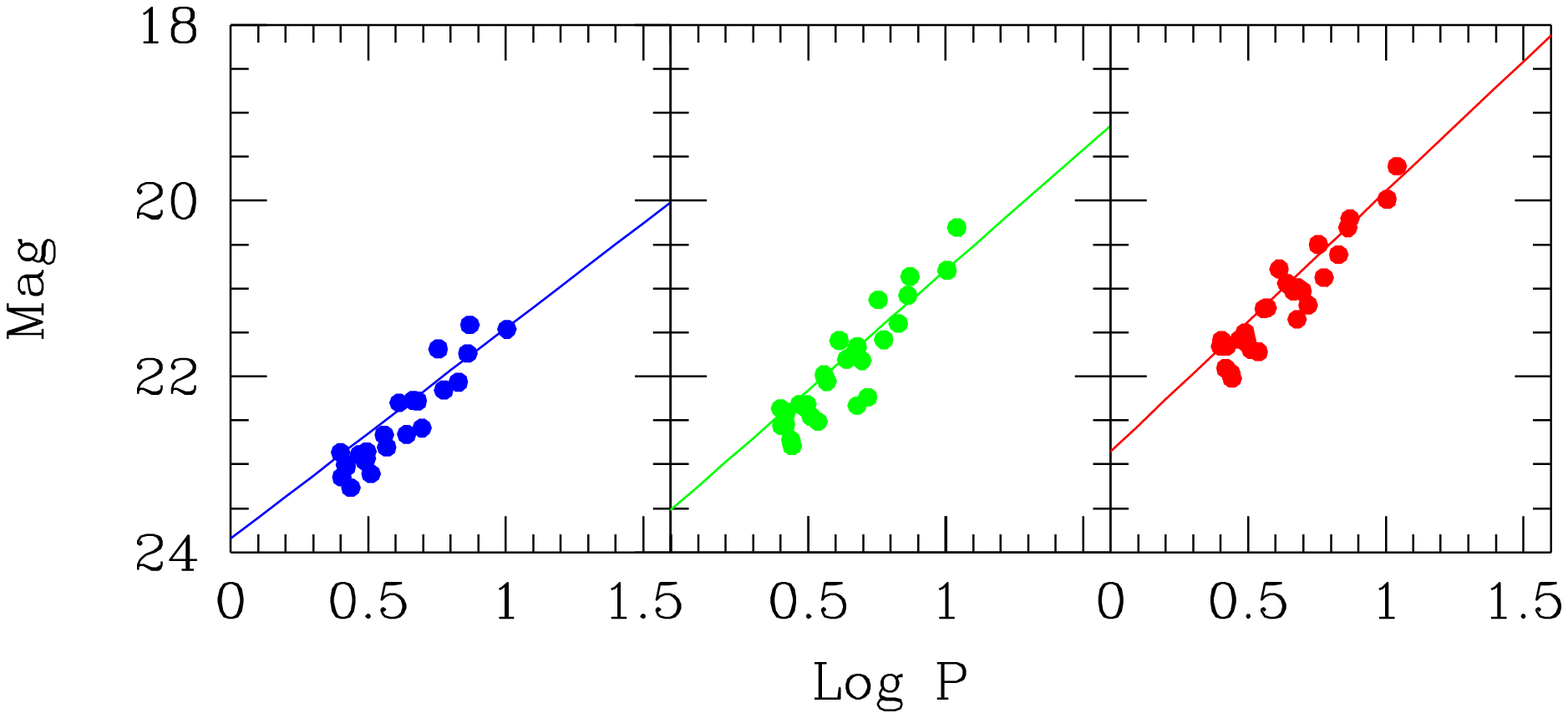} \\
\end{array}$
\caption{PL relations for the inner field (left hand panel) and the outer field (right hand panel), $B$,$V$ and $I_{C}$ go from left to right. The lines correspond to the LMC PL relations described in \citet{2007A&A...476...73F}. The magnitudes have not been corrected for reddening}
\label{bvi-pl-fig}
\end{minipage}
\end{figure*}

\begin{figure}
\includegraphics[width=84mm]{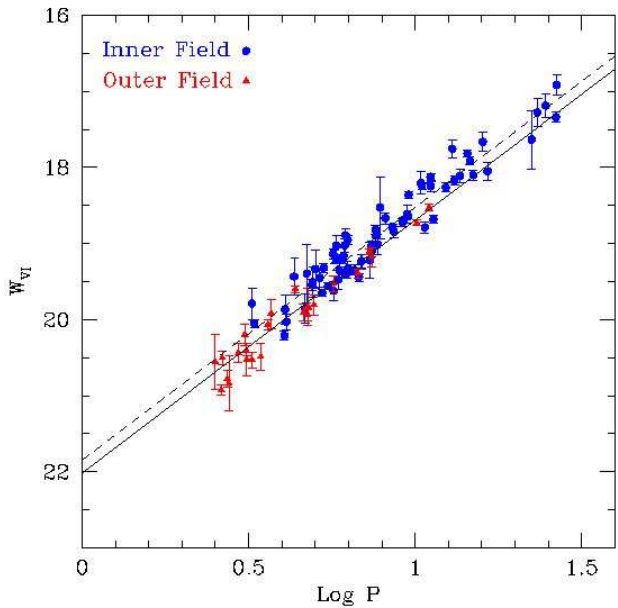} 
\caption{Reddening-free Wesenheit $W_{vi}$ PL relations for the inner field (blue circles) and the outer field (red triangles).}
\label{wvi-pl-fig}
\end{figure}

The zero-points in Table~\ref{different-pls} may at first glance appear discrepant. As they have been derived from the reddening-free $W_{vi}$ PL relations we can safely rule out an erroneous extinction correction as the source of the difference. There are three other possible explanations for the difference we observe:
\begin{enumerate}
 \item Blending affecting the magnitudes of Cepheids in the inner field.
 \item Differing period distributions of the two samples and a possible slope change at P = 10 days.
 \item Differing metallicity in the two fields.
\end{enumerate}

\subsubsection{Blending}
\label{blending}
The most obvious suggestion to explain discrepant values for the distance moduli is the fact the the inner field is extremely crowded, and there may be Cepheids with undetected companions in the sample which will affect any PL relation zero-point that we try to fit. However, we do not believe that this is the cause of the observed deviation.

Firstly, our procedure is based on {\sc allframe} which uses a deep image, made by stacking many individual exposures, to do the photometry on each individual frame. When the photometry is done on this deep reference frame, the S/N for each object is much higher than on individual images, allowing us to better recognise, and classify as such, stars that may be blended. Secondly, our comparison of the photometry with \citet{2006AJ....131.2478M} shows excellent agreement. They use images with individual exposure times of 60 sec in $BV$ and 150 sec in $I$; in other words they are up to 5 times less deep than our data.  This means that the level of crowding and blending will be less severe than in our case. Yet, our photometry is found to be in excellent agreement with theirs, for both fields, over the whole magnitude range covered by both surveys. If we were sensitive to blending levels that Massey et al. do not see, there should be a systematic offset between the two catalogues. However, as we do not see a quantifiable offset in either Figure~\ref{comparison-fig} or Table~\ref{diffs-tab} we can safely rule out blending as the cause of the difference in distance modulus for the two fields.

We can also rule out blending as the cause on a theoretical level. \citet{2000PASP..112..177F} addressed this issue, in the context of the HST Key-Project, with artificial star tests. They did this for 2 galaxies with fairly different levels of crowding. Their main conclusion was that this has little impact on distance moduli, provided the Cepheids used to derive the distance are carefully selected, as the ones in our final sample were. It is useful to note however, we found the same $\Delta \mu$ between our outer and inner fields using both the raw and selected Cepheid samples, although the PL relation constructed from the selected sample has a much lower dispersion than that of the raw sample. This reduces the uncertainty on the fit, but also provides indirect evidence that blending is not an issue in our sample.  

\subsubsection{Period Distribution and Slope Changes}
\label{period-distribtion}
\begin{figure}
\includegraphics[width=84mm]{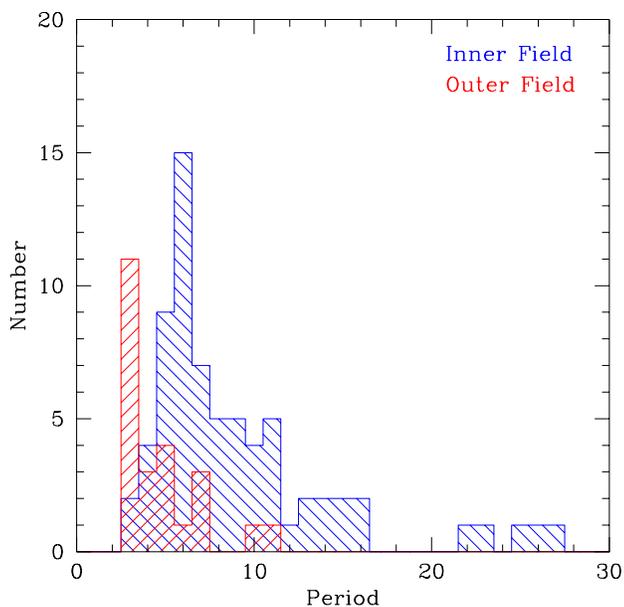} 
\caption{Period distribution of the two Cepheid samples used to fit the $W_{vi}$ PL relations}
\label{periods}
\end{figure}

The second possibility for the differing values for the distance modulus of M33 is the period range of the two samples. The distribution of periods in the final, selected sample is shown in Figure~\ref{periods}. It is quite clear that the period distribution of the two regions is not the same. This is for two reasons. Firstly, the short period Cepheids are fainter so we detect more of them in the outer field where the detection limit is lower. Secondly, long period Cepheids are rarer than their short period counterparts. Therefore, we expect the ratio of long period Cepheids to total number of stars detected to be quite low, hence they will be preferentially detected in the region with a higher number of stars; in this case the inner region.

\begin{figure}
\includegraphics[width=84mm]{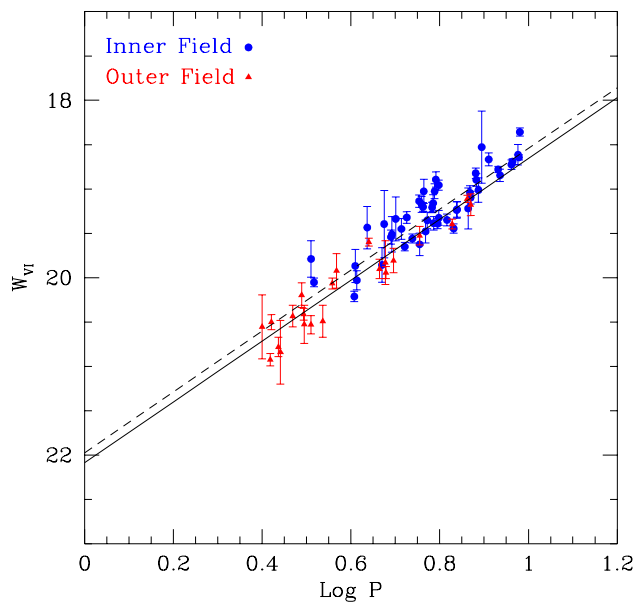} 
\caption{$W_{vi}$ PL relation fit using $\log P < 1$ criteria. The two slopes were found to be identical, however the zero-points still differ at the $3\sigma$ level.}
\label{lowP}
\end{figure}

There is some observational evidence that the slope of the $V$ and $I_C$ PL relations change around 10 days (e.g. \citet{2009APJ..693..691N}). However, there is still debate about whether this effect is real, and if the same non-linearity will appear in the $W_{vi}$ relation. To rule out any possible slope change as the cause of the discrepant distance moduli the samples were both cut to $\log P < 1$ and the PL relations refitted. Note that no attempt is made to determine if any non-linearity is present or to quantify the effect, just to eliminate the possibility; such tests would require a much larger Cepheid sample than is used in this work. As a first test both the slope and zero-point of the fits were allowed to vary. The two slopes had the same value, as expected, but again the two zero-points were different. To bring the uncertainty on the zero-points down the slope was then fixed and the zero-points refitted. The resulting fit is shown in Figure~\ref{lowP}. Using the fixed slopes gives zero-points of $21.90 \pm 0.02$ and $22.03 \pm 0.04$ for the inner and outer fields respectively. 

The fact that the zero-points are still discrepant at the $3\sigma$ level when we use a fixed period range demonstrates that this effect, whatever the cause, is universal over the period range and is not due to the possible non-linearity of the PL relation.

\subsubsection{Metallicity}
\label{metal}
As we have successfully ruled out blending and the period distribution as causes of the discrepant distance modulus the only remaining explanation is the differing chemical abundances in the two environments. The zero-points and distance moduli for each fit are shown in Table~\ref{different-pls}. 
\begin{table}
 \caption{Distance moduli from the $W_{vi}$ PL relations using the whole Cepheid sample and the period limited sample.}
\begin{center} 
 \begin{tabular}{l c c} \hline \hline
Field & $b_{LMC}$ & $\mu$ \\ \hline
 All P & &\\ \hline
Inner & $21.85 \pm 0.02$ & $24.37 \pm 0.02$ \\
Outer & $22.01 \pm 0.03$ & $24.54 \pm 0.03$  \\ \hline
 $\log P < 1$ & &\\ \hline
Inner & $21.90 \pm 0.02$ & $24.42 \pm 0.02$ \\
Outer & $22.03 \pm 0.04$ & $24.55 \pm 0.04$  \\ \hline
\end{tabular}
\label{different-pls}
\end{center}
\end{table}

\begin{figure}
\includegraphics[width=84mm]{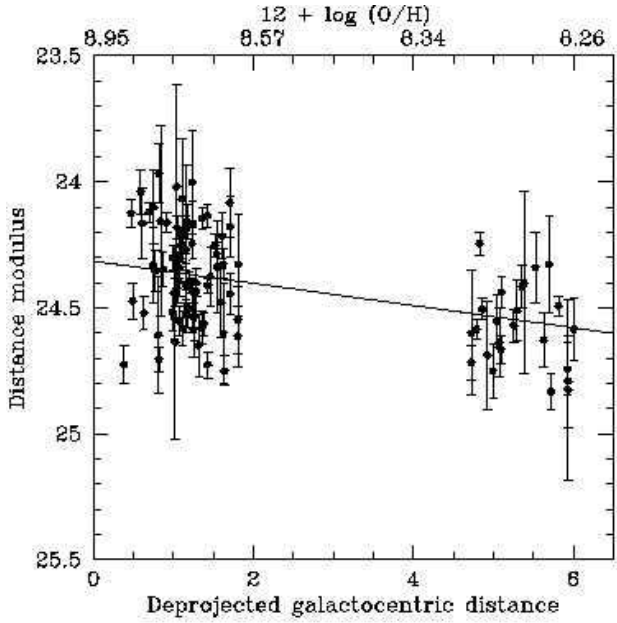} 
\caption{Measured distance moduli for each Cepheid as a function of deprojected galactocentric distance and metallicity. The line is a weighted least squares fit to the individual data points.}
\label{distances}
\end{figure}

To quantify the effect of metallicity we use $\gamma$, as defined in Equation~\ref{gamma}. We can substitute $\delta(m-M)_0$ for $\Delta \mu = 0.16 \pm 0.04$ mag, using the full sample distance moduli, and $\delta Z = -0.556 \pm 0.101 \text{ dex}$. This leads to a metallicity correction of $\gamma = -0.29 \pm 0.11$ mag dex$^{-1}$. This is compatible with recent measurements, such as that of \citet{2004ApJ...608...42S}. Our final distance measurement for M33 is calculated by assuming an LMC metallicity of $8.34$ \citep{2004ApJ...608...42S}.  Using $\gamma = -0.29 \pm 0.11$ we find that the metallicity corrected distance modulus to M33 is $\mu_{\gamma} = 24.53 \pm 0.11$, consistent with recent TRGB distance measurements. The individual distance moduli derived for each Cepheid are shown in Figure~\ref{distances}.

The uncertainty on our result is significant. By using a larger Cepheid sample the contribution from the distance moduli could be reduced. Also, the sample only covers the two extremes of the metallicity gradient. We will investigate the effect further using the data from \citet{2006MNRAS.371.1405H}. Their sample contains over 2000 Cepheids spread over the entire galaxy, and hence the entire [O/H] range. This will allow us to fully investigate the effect over a wide range of [O/H], rather than just fitting a straight line to two points, and the use of so many Cepheids will bring the random uncertainty down considerably.

We note that the largest contribution to the error budget is the uncertainty on the metallicity gradient of M33.  This could be reduced using a large sample of Cepheids. Recent work by \citet{2006ApJ...653L.101B} has shown that the two periods of double mode Cepheids can be used to calculate their metallicity. By using a large sample of these stars across the whole galaxy, in combination with observations of fundamental mode Cepheids covering the same area it would be possible to calculate $\gamma$ with a much higher degree of accuracy.

\section{Conclusions}
\label{conclusions}

We have produced a photometric catalogue in $B$,$V$ and $I_{C}$ for over 200,000 objects in two regions of M33, identifying 167 fundamental mode Cepheids. Using the reddening free Wesenheit LMC PL relations we have established distances to the two regions as $\mu_{inner} = 24.37 \pm 0.02$ mag and $\mu_{outer} = 24.53 \pm 0.03$ mag. The difference in these two distance moduli is believed to come from the effect of metallicity on the zero-point of the PL relation, and hence its effect on Cepheid distances. We have quantified this effect by the parameter $\gamma = \dfrac{\delta(m-M)_{0}}{\delta[O/H]}$ and find a value of $\gamma = -0.29 \pm 0.11$ mag dex$^{-1}$, consistent with other recent estimates. For the range of metallicities discussed here it is appropriate to include $\gamma$ in the PL relation. This leads to the numerical PL shown in Equation~\ref{altered-num-pl}, assuming an LMC distance modulus of 18.4 mag. The metallicity of the region in question is denoted by $\log [O/H]_{mes}$.

\begin{equation}
\begin{split}
W_{vi} = -3.320 (\pm 0.011) \log P + 22.01 (\pm 0.03) \\
+ 0.29 (\pm 0.11) (\log[O/H]_{LMC} - \log[O/H]_{mes})
\end{split}
\label{altered-num-pl}
\end{equation}

The analysis presented here assumes that \textit{only} the zero-point of the PL relation changes with metallicity; the slope does not significantly change between the MW and LMC metallicities. We have repeated the analysis using the MW slope of the $W_{vi}$ PL relation from F07 and we found the same value for $\gamma$. This is an important result as it means that we can find reddening- and metallicity-corrected Cepheid distances for any population from two colour ($V$ and $I_{C}$) photometry without the time consuming task of calculating internal $A_{V}$ values for other galaxies.

As noted above, the final result and particularly the error are significantly affected by the value of metallicity one uses for each region.  We have used the recent results from \citet{2007A&A...470..865M}.  \citet{2006ApJ...637..741C} found a much shallower gradient however. Using their values of metallicity in our two fields, we find $\gamma = -2.899 \pm 0.065$ mag dex$^{-1}$, incompatible with other determinations of this parameter. Such a large value of $\gamma$ would have dramatic effects on the distance scale. In particular, it would change the distance to the LMC by so much that the Cepheid distances would become incompatible with all other measurements. The difference in distance moduli between the two fields is an argument in favour of a rather large metallicity gradient in M33. However, the metallicity gradient of M33 is still subject to strong debate and, to quote \citet{2008ApJ...681..269K}, ``...the expectation of homogeneous azimuthal metallicity in patchy star-forming spiral galaxies seems naive...''. Our extension of the work of \citet{2006ApJ...653L.101B} based on double-mode Cepheids in M33 will shed light on this discrepancy by measuring the metallicity and its gradient using many more positions.  Hopefully this paper serves as an example of the need for better determinations of the metallicities of \textit{all} galaxies where Cepheids have been observed.

It is also important to note the correlation between this work and the Hubble Key Project \citep{2001ApJ...553...47F}; although the M33 Cepheids have shorter periods than the average in the HST project, the value of $\gamma$ that we find is consistent with their work. 

We will further investigate the effect of metallicity by applying the methodology described above to the catalogues from \citet{2006MNRAS.371.1405H}. With the large number of Cepheids we will be able to sample the whole [O/H] range of the galaxy and the whole $\log P - Mag$ parameter space, finally calibrating the effect that has plagued the Cepheid PL relation and distance scale for many decades.

\section{Acknowledgements}
\label{acknowlegements}

We would like to profusely thank J.-P. Beaulieu and J.-B. Marquette for the M33 fundamental mode Cepheid catalogues which have proved to be invaluable to this work. We are indebted to P.~Stetson and K.~Stanek for their assistance with creating and testing the photometry pipelines. We thank the referee, P. Fouqu\'e, for his swift and thorough report that significantly improved the content and presentation of this paper. VS is supported by an STFC studentship.

\label{references}

\end{document}